\documentclass[letterpaper, 10 pt, conference]{ieeeconf}  

\IEEEoverridecommandlockouts                              

\usepackage{graphicx} 
\usepackage{amsmath} 
\usepackage{multicol}
\usepackage{gensymb}
\usepackage[ruled,vlined]{algorithm2e}
\makeatletter
\newcommand{\removelatexerror}{\let\@latex@error\@gobble}
\makeatother
\usepackage{amsmath}
\usepackage{cases}
\usepackage{xcolor}
\usepackage{hyperref}
\hypersetup{colorlinks=true,linkcolor=blue,linktocpage}

\title{\LARGE \bf Real-time Emotion Appraisal with Circumplex Model for Human-Robot Interaction}

\author{Sarwar Hussain Paplu$^{1}$, Chinmaya Mishra$^{2}$ and Karsten Berns$^{3}$
\thanks{$^{1}$Sarwar Hussain Paplu is a PhD student at the department of Computer Science,
        Technische Universität Kaiserslautern, 67663 Kaiserslautern, Germany
        {\tt\small paplu@cs.uni-kl.de}}%
\thanks{$^{2}$Chinmaya Mishra is a MSCA PhD Fellow at Furhat Robotics, Stockholm and Max Planck Institute for Psycholinguistics, Nijmegen.
        {\tt\small Chinmaya.mishra@mpi.nl}}%
\thanks{$^{3}$Karsten Berns is a professor at the department of Computer Science and chair of Robotics Research Lab, Technische Universität Kaiserslautern, 67663 Kaiserslautern, Germany
        {\tt\small berns@informatik.uni-kl.de}}%
}

\begin{document}
\bstctlcite{IEEEexample:BSTcontrol}
\maketitle
\thispagestyle{empty}
\pagestyle{empty}

\begin{abstract}
Emotions are the intrinsic or extrinsic representations of our experiences. The importance of emotions during a human-human interaction is immense as it formulates the basis of our interaction framework. There are several approaches in psychology to evaluate emotional states in humans based on the perceived stimuli. However, the topic has been less explored as far as human-robot interaction is concerned. This paper uses an appropriate emotion appraisal mechanism from psychology, generating an emotional state in a humanoid robot on-the-fly during human-robot interaction. Since the exhibition of only six basic emotions is not sufficient to cater to diverse situations, the use of the Circumplex Model in this work has allowed the life-sized robot called ROBIN to experience 28 emotional states in different interaction scenarios. Realistic robot behaviour has been generated based on the proposed appraisal system in various interaction scenarios. 
\end{abstract}


\section{INTRODUCTION}
\label{intro}
The answer to the very fundamental question of ``Why we do what we do?'' has always been a challenging one in the field of psychology. There are intense debates as to what really motivates us to achieve our goals or to drive our behavior. A proper understanding of goals or motives can be vital in our understanding of human behavior~\cite{mietzel2005wege}. Human motivation is the experience of our desires to get something or our tendency to avoid something. The hierarchical organization of the human motivation system leads to the self-regulation of interaction and behavior~\cite{siegert2004toward}. Ordering of motives or goals based on the priority of our needs is crucial in understanding our actions, reactions or expressions~\cite{mcleod2007maslow}. In other words, human-human interaction is highly influenced by the hierarchical structure of the motivational system. It is important to know the degree of satisfaction of a motive to assess the achievement of our goals. This is known as valence. The degree or intensity of a stimulus (for example, how exciting or thrilling a stimulus is to a human) also contributes to the appraisal of emotions~\cite{ortony1988cognitive}. This is called arousal. The major aspects of experience coming out of emotional appraisal include feelings, bodily responses, expressive behaviors and sense of purpose~\cite{keltner1999functional}. 

With the advancement in the field of robotics, Human Robot Interaction (HRI) has become a focal point of research. The inclusion of robots in human environments requires a thorough understanding of the behavioral changes involved during an interaction and the robot's adaptability to various scenarios~\cite{wilcox2013optimization}. A key aspect of social robots is to perceive the interaction partner's behavior and provide a suitable affective response~\cite{breazeal2001affective}. This leads to the necessity of developing appropriate robot control architectures for the generation of behavior. The core component of such architectures comprises of a motivational and an appraisal system responsible for generating an internal emotional state for a robot.  

The existing perception system~\cite{al2016perception} of ROBIN can evaluate a large set of stimuli called ``percepts'' of an interaction partner. The reactions are mainly triggered by these visual percepts. The highest level of perception task that the robot is able to perform is the recognition of human personality traits~\cite{zafar2018real} based on non-verbal cues, making the interaction process more diverse. A large set of gestures and facial expressions have been implemented on the robot to deal with various situations during interaction. The robot's actions or reactions have been pseudo-randomized based on a given emotional state of the robot~\cite{paplu2020pseudo}, ensuring behavioural variability. However, it is observed that a rigid percept-driven interaction often leads to more of a reactive than an adaptive behavior of the robot. Therefore, there is a need for an appraisal and motivation mechanism in the robot so as to assess emotional states on-the-fly. The major focus of this paper is to evaluate interaction partners in diverse scenarios and generate an internal emotional state of a robot based on a two-dimensional (i.e., arousal and valence) appraisal mechanism. The internal emotional state is not restricted to only 6 basic emotions~\cite{ekman1999basic}. For a technical system, the realization of the world or the generation of a mental model is different in comparison with humans. In this work, we have also formalized the definitions of the dimensions of appraisal system in the context of the robot used, ensuring a robot-centered emotion appraisal. 

\section{LITERATURE SURVEY}
\label{related_work}

\begin{figure}[ht]
    \centering
    \includegraphics[width = 0.5 \textwidth]{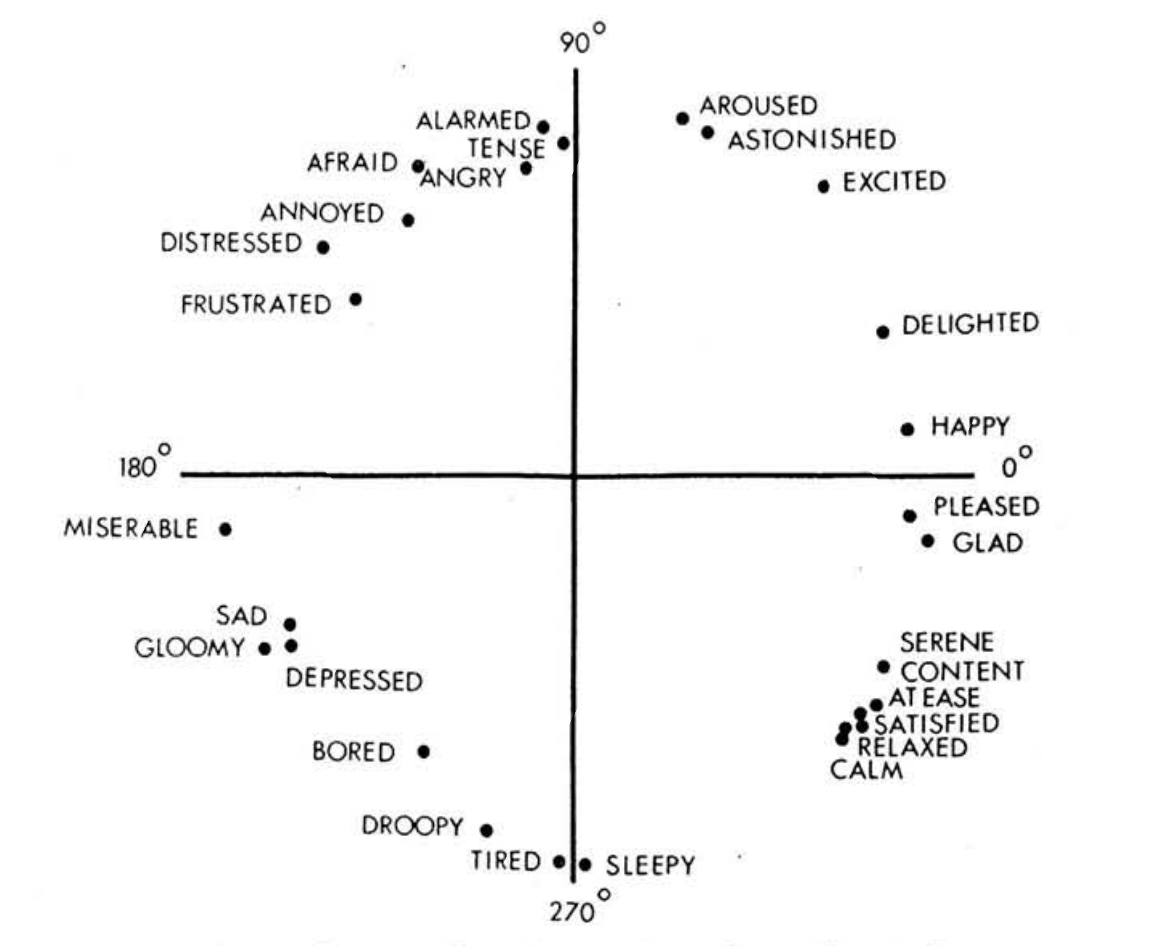}
    \caption{Circumplex model of emotions~\cite{russell1980circumplex}}
    \label{Circumplex_model}
\end{figure}

Many theories have been proposed over the decades to model emotions. Russel proposed the Circumplex Model~\cite{russell1980circumplex} in which emotions are placed on the circumference of a circle. Various human-centered experiments were conducted to prove that the placement of emotions on the circumference was correct. Fig.~\ref{Circumplex_model} shows the ordering of the emotions in a 2D space where x-axis is pleasure-displeasure i.e., valence and y-axis is the degree of arousal. 

Arousal and valence are the responses to certain stimuli presented to a group of participants. According to this model, the emotion words are not discretely separated in the 2D space, rather the points on the circumference represent the instance where the emotion is the strongest. As one point moves away towards the other, the membership of the emotion decreases from the emotion at the point of origin and increases for the emotion at the point of destination. For example, when one moves from pleased to happy, the emotion becomes~\textit{less pleased} and~\textit{more happy}. A total of 28 emotion words, after being evaluated by the participants,  found to fall meaningfully on the circle. The model proposes that the space in the middle of the circle is the ``neutral state''. The area of the circumference depends on the implementation and interpretation of the arousal and valence dimensions. 

Mehrabian \textit{et al.}~\cite{mehrabian1980basic} proposed a 3D emotion space model with \textit{pleasure}, arousal and dominance being the dimensions. Breazeal~\cite{breazeal2001affective} discussed the affective space for the proposed emotion model for a robot called KISMET. Arousal(A), Valence(V) and Stance (S)  used as the dimensions in the emotion space. Releasers are activated when a percept activates it above a certain threshold. The releasers are tagged with some affective information [A,V,S] where each tag has an associated intensity for the AVS dimensions. As a result, the emotion arbitration associates an emotion which in turn gets a specific emotion-based behavior from the behavior system. Finally, a behavior is executed in the motor systems. Hirth~\cite{Hirth12b} proposed a robot control architecture for social robots. Three dimensional appraisal system was used with arousal, valence and stance being the dimensions. With the motives or goals defined, the robot tries to achieve all the motives at any given time. The motive with the highest satisfaction is selected to influence the behavior of the robot. This approach was applied only for a gaming scenario. Moreover, the appraisal systems explained above triggers mainly six basic goal-directed emotions. In addition, the combination of speech, gesture and facial expressions to represent an affective behaviour was less explored.

To deal with the situational as well as scenario-oriented interactions, there is a need for a diverse set of emotional states and the display of appropriate behaviour based on the emotional state. The use of emotional space of the Circumplex Model~\cite{russell1980circumplex} can broaden the possibilities for a more natural and reliable interaction between a human and a robot. 

\section{ROBOT AND FRAMEWORK USED}
\label{robo_framework}

The robot used for the experiment, called ROBIN, has an ASUS Xtion Pro RGB-D Kinect sensor mounted on the chest and a built-in RGB camera on the head, forming the perception system of the robot. The robot is equipped with arms and hands, with 14 Degrees of Freedom (DoF) in each hand. A backlit projected face ensures display of various facial expressions based on facial action units. Moreover, there is a dialog  system implemented in the system, establishing interaction with humans. A C++ based robotic framework called Finroc~\cite{finroc} is employed to implement applications on the robot. This work utilizes a behavior architecture called Integrated Behavior-Based Control (iB2C)~\cite{proetzsch2010development}. The basic building block of iB2C architecture is the ``Behavior module''. A behavior module represents a single behavior in an the architecture. A key component of this architecture is the ability to combine simple behaviors to generate complex behaviors. 

\section{PROPOSED DEFINITIONS}

In order to map meaning of the perceived stimuli of the robot onto the emotion space of the Circumplex Model~\cite{russell1980circumplex}, the following definitions have been proposed:

\subsection{Arousal}

Arousal can be defined as how arousing a stimulus is to the robot. Percepts of a robot can broadly be categorised into visual, audio and physical percepts. Visual percepts comprise of anything perceived from the visual system of the robot. It may be facial expressions, body posture/ gestures, gaze, location etc. It is proposed that arousal is directly influenced by the intensity of the visual percepts and the proximity of located objects/ interlocutor. Additionally, the speed of motion influences the arousal value, e.g., a very slow motion leads to negative arousal. 
\begin{equation}
     A_v = w_1*RF + w_2*PR + w_2*SM
\end{equation}
$A_v$ is the Arousal value for visual percepts (normalized between $0$ and $1$), $RF$ stands for ``recognised features'', $PR$ is the proximity of an interlocutor, $SM$ is speed or degree of movement of the percept. $w_1, w_2 \;and\; w_3$ are the weights.
    
\subsection{Valence}

Valence is directly dependent on the satisfaction of a motive, in which a higher satisfaction value leads to a high Valence value and vice versa. It is argued that the attribute pleasant or unpleasant of any percept is dependent on the current motive/goal of the robot. 
\begin{equation}
    V = f(S)   
\end{equation}
where, $V$ denotes valence, $S$ denotes the motive's satisfaction value. 
For example, if a person does not pay attention to the conversation process with the robot, the satisfaction level of the Interact motive will gradually go down. At one stage, it reaches a threshold, triggering an unsatisfied state. This will eventually lead to a shift from this motive. 

\subsection{Gestures vs. Behaviors}
\label{gesture_behavior_def}

There is a fine line between gestures and behaviours. For the sake of clarity and consistency, we define ``gesture'' as the movement of body parts to convey a specific sentiment or message and ``behavior'' as a combination of gestures, facial expressions and speech in response to a specific stimulus.

\section{Motivation System}
\label{motivation_system}

The goals for humans' actions are represented by motives. The perceived stimuli from the perception system are fed into each ``motive'' as inputs. The output from each ``motive'' is its satisfaction value used to determine the valence in the appraisal system. The motives are implemented in a hierarchical fashion. The bottom-most motive has the highest priority, with the top-most motive having the lowest priority. Each motive inhibits the subsequent motives with low priorities. For example, the motive ``Self Preservation'', when active would inhibit the motives ``Social Motives'' and ``Self Entertainment''. 

\subsection{Motive: Obey Humans}

The motive ``Obey Human'' is responsible for incorporating the rule that the robot should obey all commands issued by the human. This rule is popularly known as Asimov's second law~\cite{anderson2008asimov} of robotics. There may always be a possibility where a human has to control the robot manually or issue commands for the robot. This motive had the highest priority and inhibits all other motives. 

\subsection{Motive: Self Preservation}

The motive ``Self Preservation'' tries to emulate the safety needs that are observed in human beings as discussed in Maslow's work~\cite{boeree2006abraham}. The safety needs vary based on the situation a person faces. However, the core idea is that there exists a need to ``find safe circumstances, stability, protection''. Taking this into account, it is proposed that there exists a need for the robot to protect and preserve itself from external harm. The robot should behave in a way that it draws attention or seek assistance from the interaction partner if the current circumstance or action poses a threat to the robot. If an interaction partner comes too close to the robot, this motive gets activated and the robot is in an unsatisfied state. The motive reaches satisfaction when the interaction partner moves to a safe distance zone. No minimum satisfaction threshold is used in this case as the robot needs to seek attention to the threat as long it exists. The robot moves away from the ``Self Preservation'' motive only when it determines that the interaction partner is at a safe distance. 

\subsection{Motive: Social Motives}
\label{social_motive}

The goal of the robot within this motive is to interact with an interaction partner and engage in a conversation. To achieve this, the motive ``Social Motives'' is split into 3 smaller motives or goals namely ``Capture Skeleton Information'', ``Greeting'' and ``Interact''. 

ROBIN needs skeleton information in order to successfully operate its perception system. Usually, the skeleton information of the interaction partner is detected very fast and does not need any manual intervention. However, at times there have been instances where the interaction partner had to move his/her hands or position himself/herself at various distances and postures in order for ROBIN to detect the skeleton. This Motive is responsible to guide the interaction partner till ROBIN acquires the skeleton information. When ROBIN detects the face of an interaction partner, with no skeleton information of the interlocutor available, the motive is activated but in an unsatisfied state. The motive reaches satisfaction when skeleton information is available. No minimum satisfaction threshold is defined in this case as the motive needed to be active as long as there is no skeleton information. 

Once the skeleton information is available, the next step is to greet. As is the case in normal human-human interaction, people usually begin their interaction with a greeting. The greeting can be a simple hand gesture, a verbal greeting or a combination of both. If an interaction partner is detected and the Greeting motive has not been activated before, this motive gets activated with a low satisfaction value. The \textit{``first time''} flag is used in our implementation to record this information. This ensures that an interaction partner is greeted only once after being identified and not multiple times, triggering more realistic interaction. Once the interaction partner greets back, the motive gains high score on satisfaction. As ROBIN does not have the capability to utilize any audio information, the ``greeting back'' gesture is realized based on the recognition of a set of relevant hand gestures.

The motive ``Interact'' is responsible to make interaction between the robot and the interaction partner possible. The goal of the robot withing this motive is to get engaged with the interaction partner and display behaviors in a natural human-like manner. When there is an interaction partner available, the motive gets activated but with an unsatisfied state. To determine if the human is interested in the interaction, the perception system observes if the person is attentive and looking forward. In case, the interlocutor looks away or looks back or does anything suggesting that he/she is not interested, the Satisfaction value decreases by ``Neg\_Step''. The Satisfaction value increases by ``Pos\_Step'' when the person seems interested to interact. The values of ``Pos\_Step'' and ``Neg\_Step'' are used to control the rate by which the Satisfaction value changed over time, and the above mentioned values are experimentally found to keep the interaction natural. This is shown in table~\ref{interact_par}.
 
\begin{table}[ht]
\caption{Parameters for Motive Interact}
\label{interact_par}
\begin{center}
\begin{tabular}{|p{3.9cm}|p{3cm}|}
\hline
\textbf{Property} & \textbf{Value} \\
\hline
Motive Type & Event Based \\
\hline
Triggering Events & Human present  \\
\hline
Satisfying Events & Looking forward \\
\hline
Maximum Satisfaction Threshold & 0.9 \\
\hline
Minimum Satisfaction Threshold & -0.8 \\
\hline
Pos\_Step & 0.003 \\
\hline
Neg\_Step & -0.02 \\
\hline
\end{tabular}
\end{center}
\end{table}

\subsection{Motive: Self Entertainment}
\label{self_entertainment}

This motive is responsible to engage the robot in random activities, provided there is a lack of perceptual stimuli. By performing various activities such as singing or acting, it is attempted to emulate the behavior of self entertainment often observed in humans. In this way, the robot is also able to attract attention of any potential interaction partners in the vicinity and thereby gain a chance to interact. The moment a human is detected, the motive switches from very low to full satisfaction. 

\subsection{Fusion of Motives}
\label{fusion_motive}

The next step is to integrate the implemented motives in a hierarchical fashion. A fusion behavior module from the iB2C architecture is used to cater the fusion. All the ``inhibition'', ``target rating'' and ``output'' signals are fed into the fusion behavior. The final output is chosen based on the \textit{``winner takes it all''} principle and only one set of output signal is passed on to the Emotion Appraisal system. The fusion behavior is responsible for maintaining the hierarchical architecture of all the motives discussed. It filters out the output signals from inactive motives and allow only the output signals from the active motive to go through.


\section{EMOTION APPRAISAL}
\label{appraisal_system}

\subsection{Perception of Stimuli}
\label{perception_system}

OpenNI and NiTE libraries enable us to detect humans by utilizing depth and Infra Red (IR) sensor of ASUS Xtion. The algorithms not only detect humans but also track them efficiently. They extract human skeleton joints using NiTE library and convert them into angles. Feature vectors are generated using angles between each joint and classify them with Support Vector Machines (SVMs). The system uses low-level perception features to understand high-level perception behaviors, e.g., head gesture recognition~\cite{saleh2015nonverbal}, facial expression recognition~\cite{al2016action}, body posture~\cite{zafar2018real} etc. These nonverbal features lie in low-level perception and can be used to recognize high-level perception behavior, when analyzed over a period of time. Movements performed by the limbs of a human play an important role in the recognition of activity. 

\subsection{Calculation of Arousal}
\label{calc_arousal}

We tagged each percept (facial expression, hand gestures, head gestures and body postures) with an intensity value ranging from 0 to 1, with 0 being the lowest intensity value and 1 being the highest. The values are set empirically. The intensity values of the perceived stimuli vary based upon the proximity from the robot and the activity or movement associated with the stimuli. For example, waving with one hand has lower stimulus intensity than waving with both hands. The distance zones proposed by E.T. Hall~\cite{hall1910hidden} have been used. A variable \textit{zone intensity} is defined and set to 1. If a person is in the social zone, the intensity of the perceived stimulus is directly reflected as the overall intensity, whereas if the person is in the personal zone, the intensity of the stimulus has 50\% weight and the \textit{zone intensity} has 50\%. If a person is in the intimate space, the stimulus intensity has 0\% weight and \textit{zone intensity} has 100\%. For public zone, the stimuli intensity has a weight of 25\% and the \textit{zone intensity} is set to 0 in public zone. The algorithm~\ref{algoArousal} have been applied to calculate arousal, where \textit{step} is set to 0.25 and \textit{weight} is set to 1. \textit{Step} value increases the level of arousal. $Weight$ indicates how fast the arousal value increases.

\begin{figure}[!ht]
\removelatexerror
\begin{algorithm}[H]
	\SetAlgoLined
	\KwResult{Arousal($A_t$)}
	\uIf{$change\;in(overall\;intensity) = 0$}
	{$A_t\;=\;A_{t-1}\;-\;step;\;$}
	\Else
	{$A_t = weight\;\cdot\;overall\;intensity$}
	limit $A_t$ to the range of [-1,1];
	\caption{Calculation of Arousal}
	\label{algoArousal}
\end{algorithm}
\end{figure}

\subsection{Calculation of Valence}

The satisfaction of a motive is calculated based on the triggering and satisfying events for a motive, where $pos\;step$ and $neg\;step$ determine the rate at which the satisfaction value increases or decreases. These values need to be be selected based on the motive in question. The algorithm \ref{algoValance} is used to calculate the valance($V$) for a motive.

\begin{figure}[!ht]
\removelatexerror
\begin{algorithm}[H]
\SetAlgoLined
\KwResult{Valance(V)}
 $V_{t-1}= 0$\;
 {$S = Satisfaction \;Value \;of \;the\;active\;motive$\;}
 $step = min(|S - V_{t-1}|, weight)$\;
 \uIf{$S > V{t-1}$}
 {$V_t = V_{t-1} + step;$}
 \Else{$V_t = V_{t-1} - step;$}
 limit $V_t$ to the range of [-1,1];
 \caption{Calculation of Valance(V)}
 \label{algoValance}
\end{algorithm}
\end{figure}

$weight$ is the maximum rate by which the Valence($V$) value changes and $step$ is used to increase or decrease the Valence value in a step-wise manner. Two threshold values, $S_{max}$ and $S_{min}$, of \textit{Satisfaction} are used to determine the activity of a motive. $S_{max}$ denotes the maximum \textit{Satisfaction} value of the motive at which the motive is satisfied and should in turn switch to being inactive. Similarly, $S_{min}$ denotes the minimum \textit{Satisfaction} value of the motive to decide the activity.
\begin{subnumcases}{a=}
   1 & $S_{min} < S < S_{max}$ \label{positive-subnum}
   \\
   0 & $otherwise$ \label{negative-subnum}
\end{subnumcases} 

\subsection{Emotion from Arousal \& Valence}
\label{emo_from_AV}

After the values for \textit{arousal} and \textit{valence} are calculated, they are used to determine an \textit{emotion state} of the robot. The 28 emotion words from Russel's work~\cite{russell1980circumplex} are used as \textit{emotion State} in this work. The emotion words fall meaningfully on the circle with the following degree values:

\begin{multicols}{2}
\begin{itemize}
    \item happy : 7.8\degree
    \item delighted : 24.9\degree
    \item Excited : 48.6\degree
    \item Astonished : 69.8\degree
    \item Aroused : 73.8\degree
    \item Tense : 92.8\degree
    \item Alarmed : 96.5\degree
    \item Angry : 99\degree
    \item Afraid : 116\degree
    \item Annoyed : 123\degree
    \item Distressed : 138\degree
    \item Frustrated : 141\degree
    \item Miserable : 188.7\degree
    \item Sad : 207.5\degree
    \item Gloomy : 209\degree
    \item Depressed : 211\degree
    \item Bored : 242\degree
    \item Droopy : 256.7\degree
    \item Tired : 267.7\degree
    \item Sleepy : 271.9\degree
    \item Calm : 316.2\degree
    \item Relaxed : 318\degree
    \item Satisfied : 319\degree
    \item At ease : 321\degree
    \item Content : 323\degree
    \item Serene : 328.6\degree
    \item Glad : 349\degree
    \item Pleased : 353.2\degree    
\end{itemize}
\end{multicols}

Emotion Space consists of two dimensions with \textit{arousal} and \textit{valence} being the dimensions. The X-axis and Y-axis represent the \textit{valence} and \textit{arousal} values respectively. The range for both the axes is $[-1,+1]$. The membership of the \textit{emotion states} are defined as sectors in an unit circle. As described by Russel, the specific degree values represented the points in the ``emotion space'' where the membership of the emotion word is maximum. Taking this as the basis, the specific degree value is taken as the midpoint of the arc of the sector for each \textit{emotion state}. For example, ``happy'' has a degree value of 7.8\degree. So, 7.8\degree \;is taken as the midpoint of an arc of the sector. The end points of the arc are calculated as the mid-point between the degree values of ``Pleased-Happy'' and ``Happy-Delighted'' respectively which are 0.5\degree\; and 16.35\degree. To calculate $\theta$ in the AV 2D space, the following function is used to convert the 2D coordinate point into degrees.
\begin{equation}
    \theta = arctan2(\frac{y}{x})
\end{equation}
$y$ and $x$ are the \textit{arousal} and \textit{valence} values respectively. Depending on the $\theta$ value obtained, the corresponding \textit{emotion state} is calculated. For example, any point in AV coordinate system that resulted in a $\theta$ value between 0.5\degree\; to 16.35\degree\; is assigned an \textit{emotion state} of ``happy''.

\section{EXPERIMENTATION \& EVALUATION }

Based on the emotion derived from the proposed appraisal system, robot behavior is generated in the form of gestures, facial expressions and dialogues. Separate lists of behaviours, comprising of these three channels, have been created and integrated with the existing XML-based dialog system of the robot. There is a direct mapping of emotional state of the robot to its relevant behaviour, enabling more autonomy to the interaction process. Human-centred evaluation of the system developed has been conducted to verify if the emotional states generated by the robot in interaction scenarios are realistic. 

\begin{figure}[ht]
    \centering
    \includegraphics[width = 0.48 \textwidth]{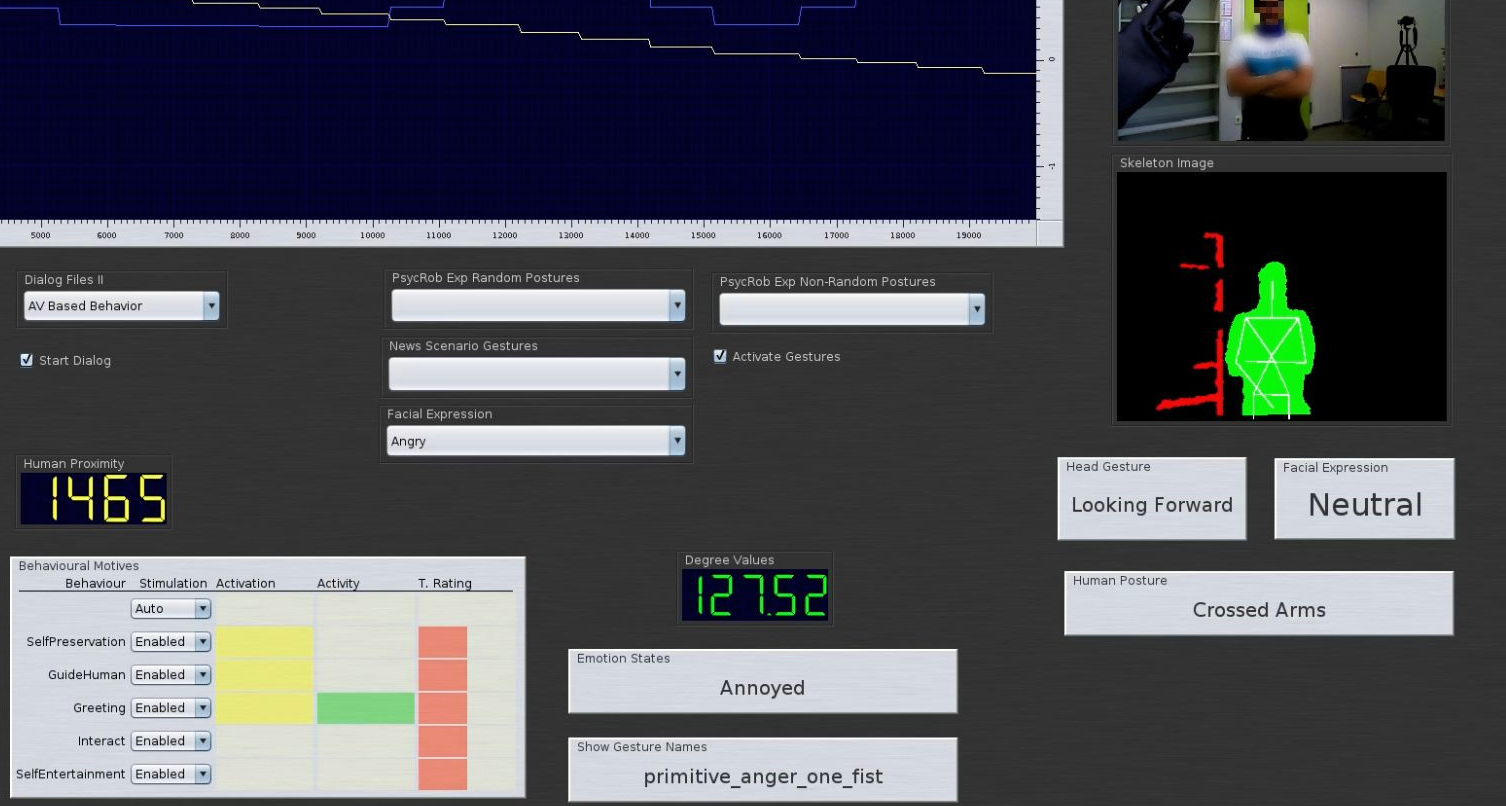}
    \caption{User interface showing existing emotional state of ROBIN}
    \label{exp_fingui}
\end{figure}

It is also important to investigate how the robot switches its motives during an interaction with humans. In a typical scenario, the robot starts greeting an interaction partner once he/she is visible by the robot. In this case, the goal or the motive of the robot is to get a greeting back from the interlocutor. The moment the robot gets a satisfying event, the valence goes high and there is a possible switch in its motive. Given this scenario, the robot ends up being in ``interact'' motive as the motive is satisfied but there is a gradual change in emotional states as the values for arousal and valence change over time based on the stimuli perceived. The robot's emotional state at a specific time of interaction can be observed in the fig.~\ref{exp_fingui}. At this point, the degree value calculated out of arousal and valence is 127.52\degree\; which triggers an emotional state of annoyance. Additionally, an ``engagement'' scenario has been created in which the robot observes if an interlocutor is paying attention to the responses or queries generated by the robot. The events that slowly trigger a transition from this motive to the other relevant motive are the actions of a human that doesn't imply engagement, for example, looking away, looking down, showing little or no physical activity etc. This motive-driven interaction process ensures much more autonomy in the robot's behaviour as compared to a reactive process in which emotional state is pre-defined. 

A total of 16 participants, university students and employees, are invited to evaluate the system. We briefly explained them the circumplex model that evaluates emotions. They are asked to interact with the robot standing in a room and observe carefully how the robot changes its internal state on its own, being in various motives. Each interaction partner was exposed to the scenarios explained earlier. The entire interaction goes on between the participant and the robot, with the system interface's screencast switched on. A snapshot of the interface can be seen in fig.~\ref{exp_fingui}. The interlocutors could see the screencast of our interface right after the interactions. The screencasting helps participants and other observers to judge if the emotional state is realistic, given the scenarios. Each of the subject is provided with a questionnaire comprising of 5 questions. The questions include: (i) is the change in robot's emotional state meaningful or realistic, (ii) is the switch between motives appropriate? (iii)  was the robot's speech, gesture and facial expressions synchronized properly? (iv) can the implemented appraisal system, in reality, comply with the circumplex model? Each of these questions had 3 options to choose: realistic, unrealistic and unclear. Additionally, there was an open-ended question, asking about the overall user experience with the interaction scenarios. 

\begin{figure}[ht]
    \centering
    \includegraphics[width = 0.48 \textwidth]{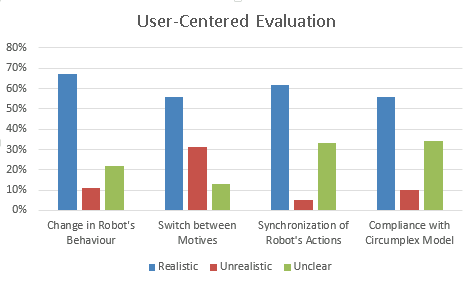}
    \caption{Questionnaire-based evaluation: percentage of users vs. various interaction aspects}
    \label{eval_res}
\end{figure}

User experience with the system implemented has been collected from the questionnaire. Fig.~\ref{eval_res} depicts a summary of the data on user-experience collected after the experiments. It can be observed that $67\%$ of the participants considered the change in the robot's behaviour during interaction to be realistic. However, the synchronization of speech, gesture and facial expressions was found out to be unrealistic by a big chunk of users ($33\%$). System latency often leads to this problem. In addition, it was unclear for $31\%$ of the interaction partners if the system complies with the Circumplex Model of psychology. The participants often expected the robot to change its motive much faster. In contrast, the proposed approach applies gradual increase or decrease in the calculation of valence. This often resulted in unexpected delay in the switch between the motives of the robot, as far as human's perspective is concerned. Overall, most of the participants expressed their satisfaction over the technical system generating emotional states and displaying relevant behaviour under some conditions.

\section{CONCLUSIONS \& FUTURE WORK }

In order to make a robot emotionally intelligent, emotion appraisal mechanism is vital. Manually informing the robot about an emotional state is an obstacle as far as intelligent human-robot interaction is concerned. This work ensures that the robot itself creates a mental model of the interaction partner and derives an emotional state on-the-fly. Experimental results showed that the robot, in most cases, managed to generate a suitable emotional state on its own based on the appraisal mechanism proposed in this work. The hyper-parameters used during the calculation of an emotional state can be fine tuned with additional experiments. Existing gestures, postures and facial expressions of the robot can be enriched to ensure better display of emotions on the robot.

\addtolength{\textheight}{-12cm}   


\bibliographystyle{IEEEtran}
\bibliography{IEEEabrv,IEEEexample}

\end{document}